\documentclass{emulateapj}
\usepackage{amsmath}
\usepackage{amssymb}
\usepackage{graphicx}

\newcommand\bmath[1] {\mbox{\boldmath$\rm #1$}}

\usepackage{natbib}
\newcommand\cf{{cf.}} 
\newcommand\eg{{e.g.}} 
\newcommand\ie{{i.e.}} 

\newcommand\K{{\rm K}} 
\newcommand\yr{{\rm yr}} 
\newcommand\Myr{{\rm M}\yr} 
\newcommand\m{{\rm m}} 
\newcommand\cm{{\rm c}\m} 
\newcommand\g{{\rm g}} 

\renewcommand\d{{\rm d}}

\begin{document}

\title{Oscillating Starless Cores: The Nonlinear Regime}

\author{Avery E. Broderick\altaffilmark{1}, Eric Keto\altaffilmark{2},
Charles J. Lada\altaffilmark{3}, Ramesh Narayan\altaffilmark{4}}
\affil{Harvard-Smithsonian Center for Astrophysics, 60 Garden
  Street, Cambridge, MA 02138, USA}
\altaffiltext{1}{abroderick@cfa.harvard.edu}
\altaffiltext{2}{keto@cfa.harvard.edu}
\altaffiltext{3}{clada@cfa.harvard.edu}
\altaffiltext{4}{rnarayan@cfa.harvard.edu}

\shorttitle{OSCILLATING STARLESS CORES: THE NONLINEAR REGIME}
\shortauthors{BRODERICK ET AL.}

\begin{abstract}
In a previous paper, we modeled the oscillations of a
thermally-supported (Bonnor-Ebert) sphere as non-radial, linear
perturbations following a standard analysis developed for stellar
pulsations. The predicted column density variations and molecular
spectral line profiles are similar to those observed in the Bok
globule B68 suggesting that the motions in some starless cores may
be oscillating perturbations on a thermally supported equilibrium structure. 
However, the linear analysis is unable to address several questions, among
them the stability, and  lifetime of the perturbations. 
In this paper we simulate the oscillations using a three-dimensional numerical 
hydrodynamic code.  We find that the oscillations are damped
predominantly by non-linear mode-coupling, and the damping time scale is
typically many oscillation periods, corresponding to a few million
years, and persisting over the inferred lifetime of gobules.
\end{abstract}


\maketitle

\section{Introduction}

Recent observations  suggest that the internal structure of some of
the small isolated molecular clouds known as starless cores or Bok
globules \citep{Bok1948} is well approximated by a balance of thermal and
gravitational forces on which is superposed a pattern of non-radial
oscillations
\citep{Lada2003}. Previously,
the internal oscillations of these starless cores have been noted
in simulations of collapse \citep{Hennebelle2003}, magnetized clouds
\citep{Galli2005}, modeled as one-dimensional radial oscillations by
means of a numerical simulation that includes radiative energy losses
\citep{KetoField2005} and as three-dimensional non-radial
perturbations assuming isothermal conditions
\citep{Keto2006}. Analysis of these models and comparison of
simulated molecular spectral line profiles with those observed in the
dark clouds L1544  \citep{Caselli2002a} and B68 \citep{Lada2003}
indicate that the motions in some dark clouds are consistent with
large-scale hydrodynamic oscillations, i.e., sound waves.
Nevertheless, a number of questions remain as yet unanswered. 

Is the underlying assumption of the oscillations as perturbations valid for
the amplitudes required to produce the observed velocities?
Based on the observations of B68, the perturbation amplitudes in
our previous three-dimensional model were $\sim25$\%. 
Does the linear model correctly
predict the velocity field for such large amplitudes
or would significant differences appear in the
non-linear regime?

What is the lifetime of the perturbations and thus the velocities
within the clouds?  For hydrodynamic oscillations to remain a viable
explanation of the velocity field observed in B68 they must have
lifetimes in excess of a crossing time.  Supersonic oscillations are strongly
damped via shocks. However, in the subsonic regime the dominant
damping mechanism is less clear.  In the one-dimensional model \citep{KetoField2005}, the
dissipation of the oscillations occurs via radiative losses with an
estimated timescale of $10\,\Myr$.  In the
non-radial perturbative model \citep{Keto2006} damping is explicitly ignored.
For mode amplitudes larger than linear perturbations, 
non-linear mode-mode coupling provides an additional
potentially significant damping mechanism.

Are the oscillations stable or growing? Are there preferred modes?
In general we expect the oscillations to damp over time, and we expect
that the longest-lived modes of oscillation would be the long wavelength 
modes. Neither the one-dimensional non-linear analysis nor the three-dimensional
linear analysis is able to adequately address this question.

In this paper we address these questions by modeling the oscillations of dark clouds
with a three-dimensional, non-linear,
numerical hydrodynamic simulation.
We find that the dominant damping mechanism is
indeed non-linear mode coupling, the longest wavelength mode lifetimes
are on the order of a few $\Myr$ and produce velocity and density
fields qualitatively similar to those observed.  Section \ref{CM}
summarizes the numerical methods employed, sections \ref{QP} and
\ref{DP} discuss the viability of quadrupole and dipole oscillation
models, and the conclusions are contained in section \ref{C}.

\section{Computational Modeling} \label{CM}

\begin{figure}[tb]
\begin{center}
\includegraphics[width=\columnwidth]{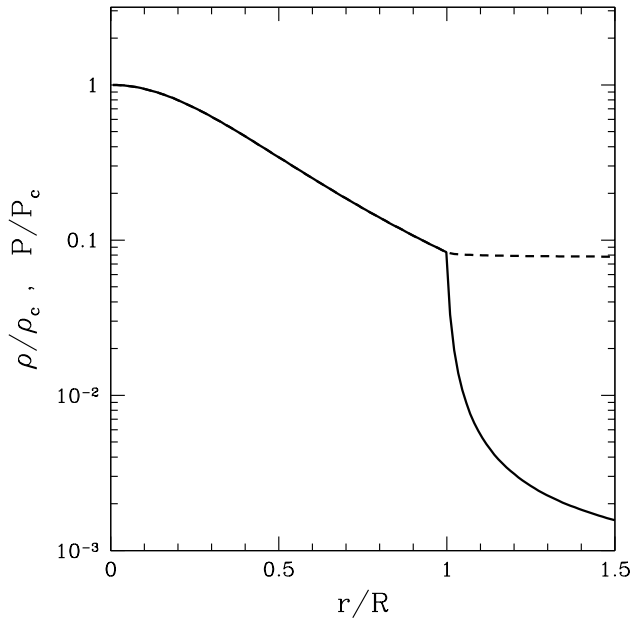}
\caption{The density (solid) and pressure (dashed) profiles of the
  unperturbed isothermal cloud (Bonnor-Ebert sphere).  Near the
  surface of the cloud ($r/R=1$) the density rapidly decreases.
  However, the transition region in the equation of state softens this
  drop over many grid zones to avoid difficulties in resolving the
  pressure gradients.} \label{fig:profiles}
\end{center}
\end{figure}

\begin{figure}[tb]
\begin{center}
\includegraphics[width=\columnwidth]{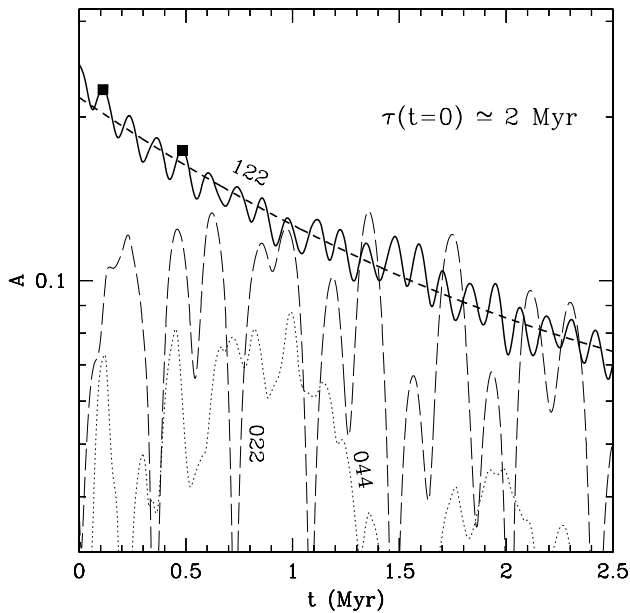}
\caption{The dimensionless amplitude of the first quadrupole p-mode,
  the (1,2,2) mode.
  The dashed line is a quadratic fit to the logarithm of the
  amplitude, with an e-folding time of roughly $2\,\Myr$ at $t=0$.
  The squares indicate the times for which the frames in Figures
  \ref{fig:q} were computed.
  The dotted and long-dashed lines show the only other two significant
  long-wavelength oscillations, listed by their $nlm$.} \label{fig:quad_amp}
\end{center}
\end{figure}

\begin{figure*}[t!]
\begin{center}
\begin{tabular}{cc}
\includegraphics[width=0.48\textwidth]{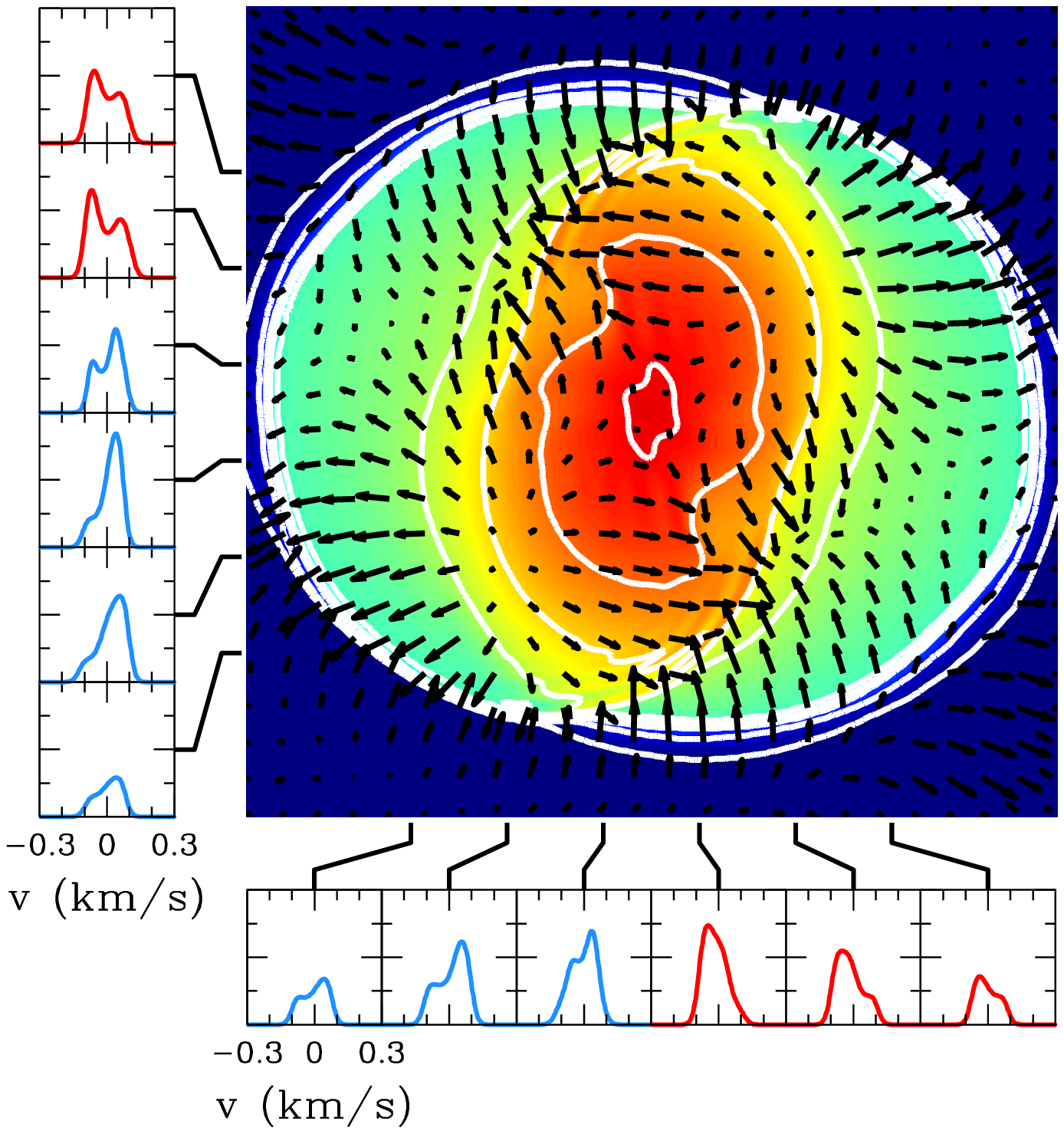}
&
\includegraphics[width=0.48\textwidth]{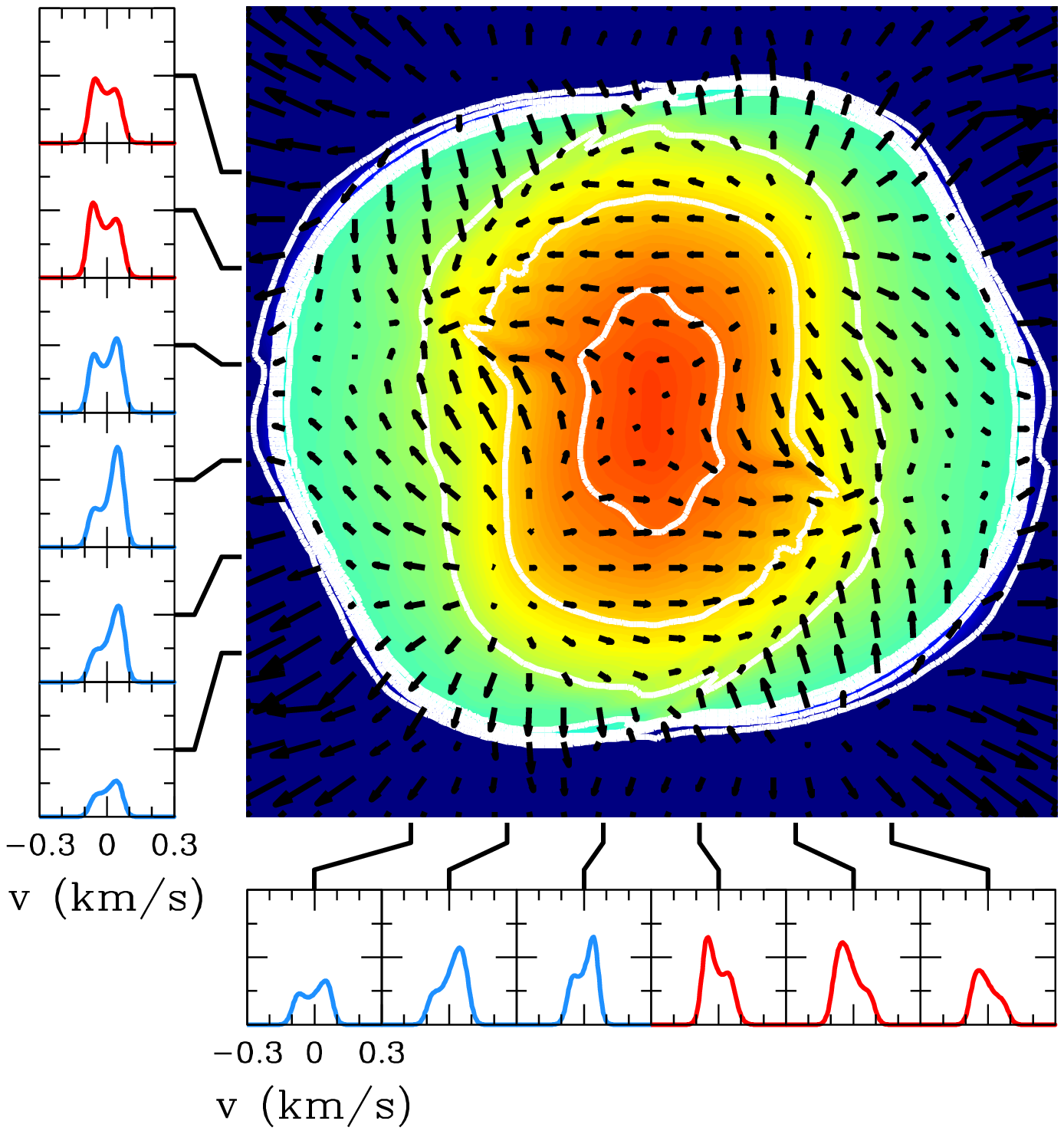}
\end{tabular}
\end{center}
\caption{The equatorial gas density, velocity, and CS(2-1) line
  profiles of the quadrupolar mode after the first half-period, (left),
  and the second full period, (right).  These correspond to the times
  marked by black squares in Figure \ref{fig:quad_amp}. Despite having
  oscillated for nearly
  $5\times10^{5}\,\yr$, the quadrupolar structure is still clearly
  evident in the density \& velocity fields as well as in the spectral
  line profiles.  Spectral line profiles colored blue (red) indicate
  approaching (retreating) gas on the near side relative to the
  observer.} \label{fig:q}
\end{figure*}

\begin{figure}[tbh!]
\begin{center}
\includegraphics*[width=\columnwidth]{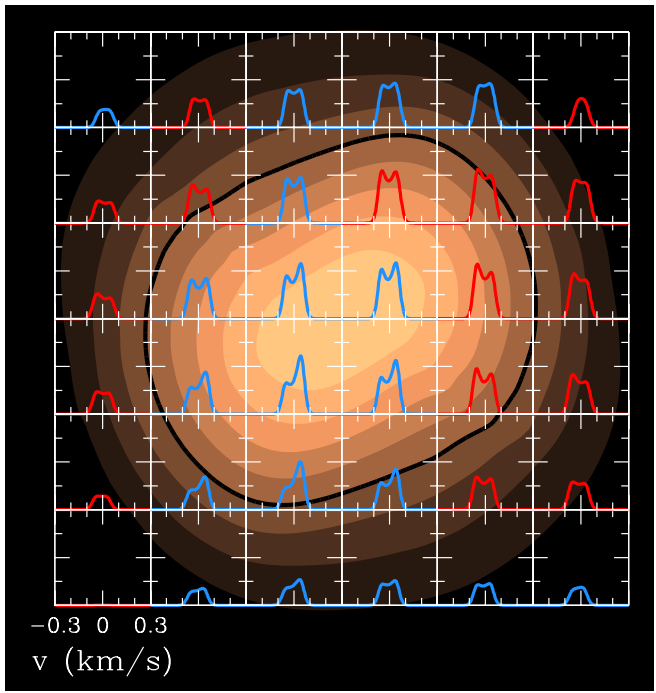}
\end{center}
\caption{A map of the CS(2-1) line profile overlaid upon the gas column
  density for the quadrupolar oscillation after nearly
  $5\times10^5\,\yr$.  This is viewed from $\lambda=30^\circ$, $\theta=30^\circ$, for
  which the pattern of spectral line profiles is similar to the Bok
  globule B68 (\cf~Figure 7 of \citealt{Keto2006}).  The
  gas density is shown as a filled contour map, with contours equally
  spaced in $10\%$ increments of the maximum value and the $50\%$
  contour is marked by the thick black line.  The CS(2-1) line profiles
  are associated with lines of sight centered in each subplot.  Blue (red)
  lines correspond to approaching (retreating) gas on the nearside of
  the cloud relative to the observer.  Note the prominent double reversal of the
  line-of-sight motions and the substantial distortion of
  the cloud shape, as seen in B68 (and in the same relative
  orientation).}\label{fig:qcds} 
\end{figure}

\begin{figure}[tb]
\begin{center}
\includegraphics[width=\columnwidth]{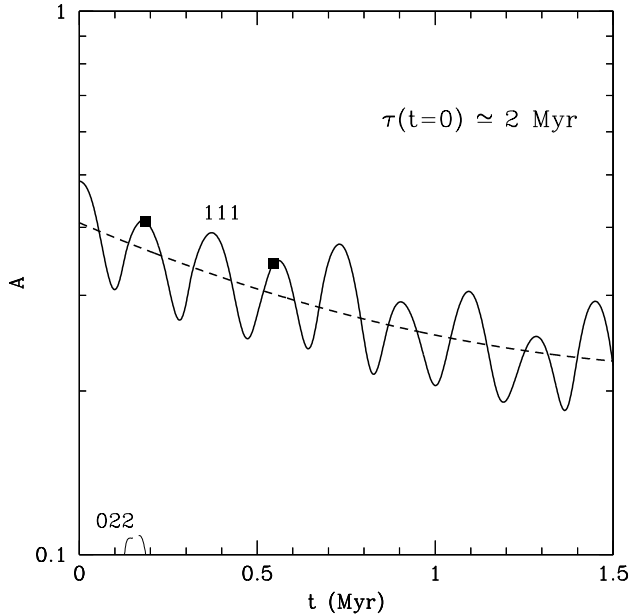}
\caption{The dimensionless amplitude of the first dipole p-mode, the
  (1,1,1) mode.
  The dashed line is a quadratic fit to the logarithm of the
  amplitude, with an e-folding time of roughly $2\,\Myr$ at $t=0$.
  The squares indicate the times for which the frames in Figure
  \ref{fig:d} were computed.
  The long-dashed line shows the $(0,2,2)$ mode, the only other significant
  oscillation.} \label{fig:dip_amp}
\end{center}
\end{figure}

\begin{figure*}[tbh]
\begin{center}
\begin{tabular}{cc}
\includegraphics[width=0.48\textwidth]{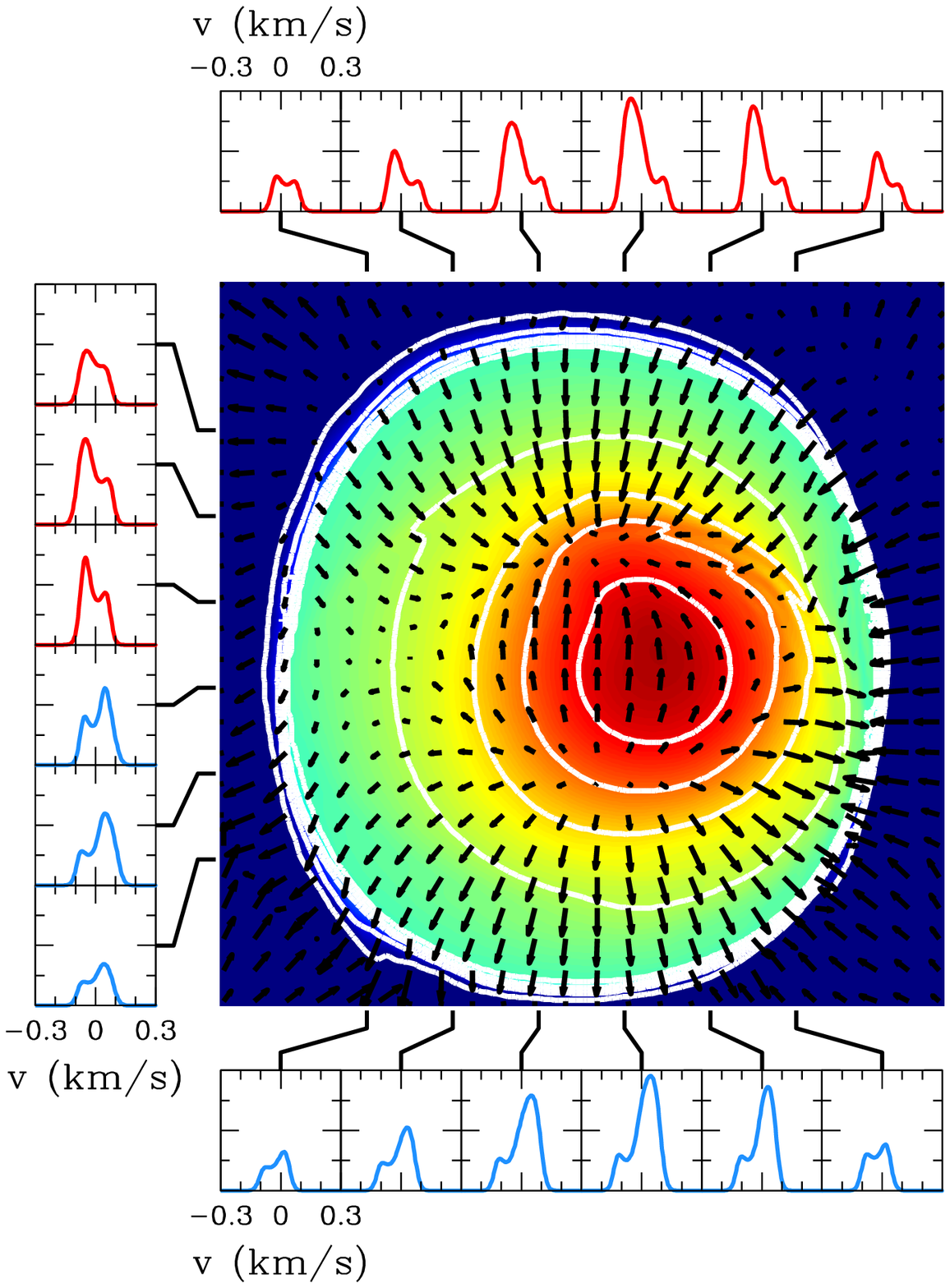}
&
\includegraphics[width=0.48\textwidth]{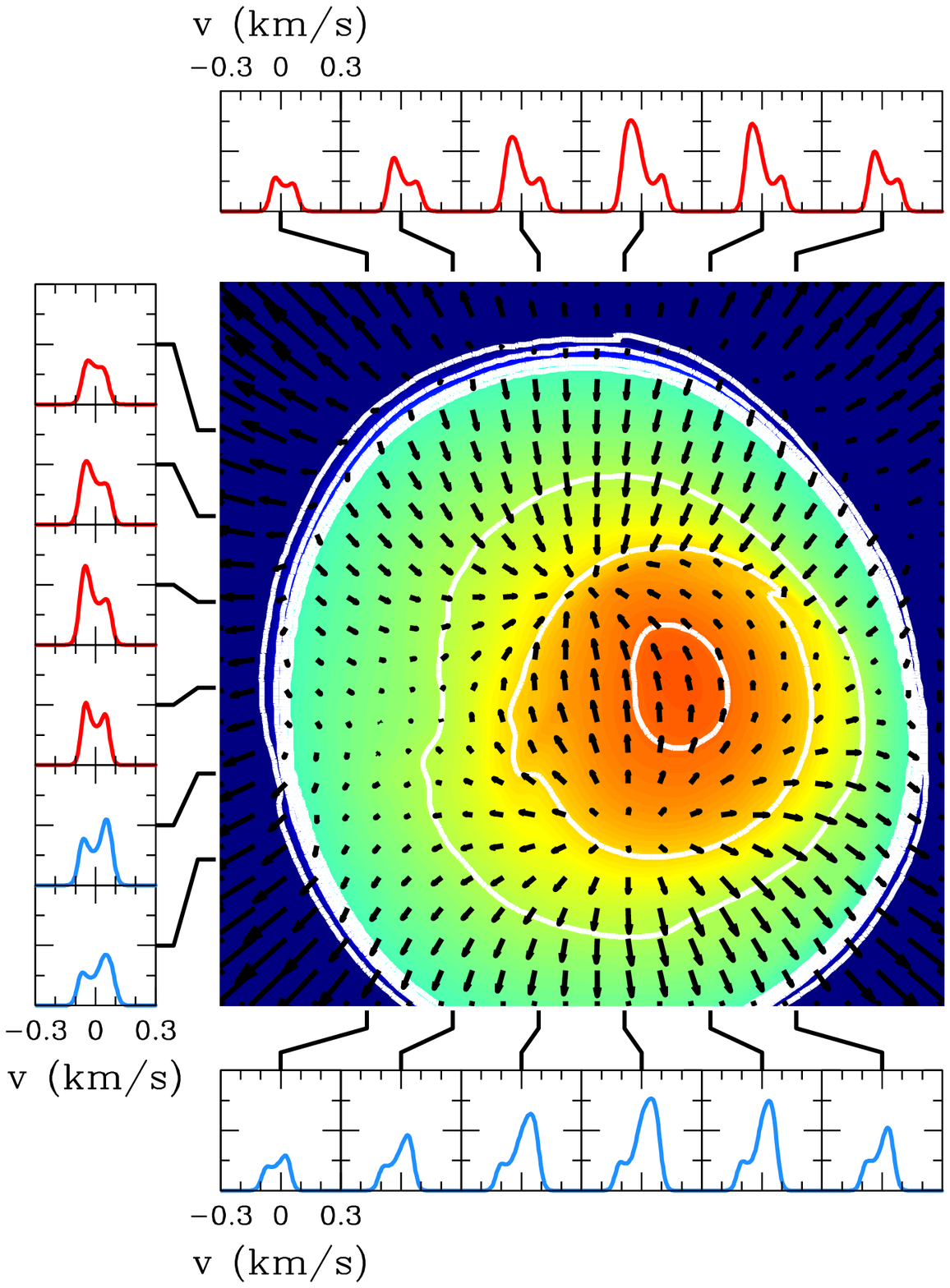}
\\
(a) & (b)
\end{tabular}
\end{center}
\caption{The equatorial gas density, velocity, and CS(2-1) line
  profiles of the dipole mode after the first half-period, (a),
  and the second full period, (b).  These correspond to the times
  marked by black squares in Figure \ref{fig:dip_amp}.  Despite having
  oscillated for more than
  $5\times10^{5}\,\yr$, the dipolar structure is still clearly
  evident in the density \& velocity fields as well as in the spectral
  line profiles.  Spectral line profiles colored blue (red) indicate
  approaching (retreating) gas on the near side relative to the
  observer.  Note that the spectra along the bottom would seem to
  imply expansion while those along the top would imply collapse.
  Furthermore, spectra from the side suggests rotation.
  Thus inferring the global state of the cloud from line profiles may
  be complicated by the presence of large-scale
  oscillations.}\label{fig:d}
\end{figure*}

\begin{figure}[tbh!]
\begin{center}
\includegraphics*[width=\columnwidth]{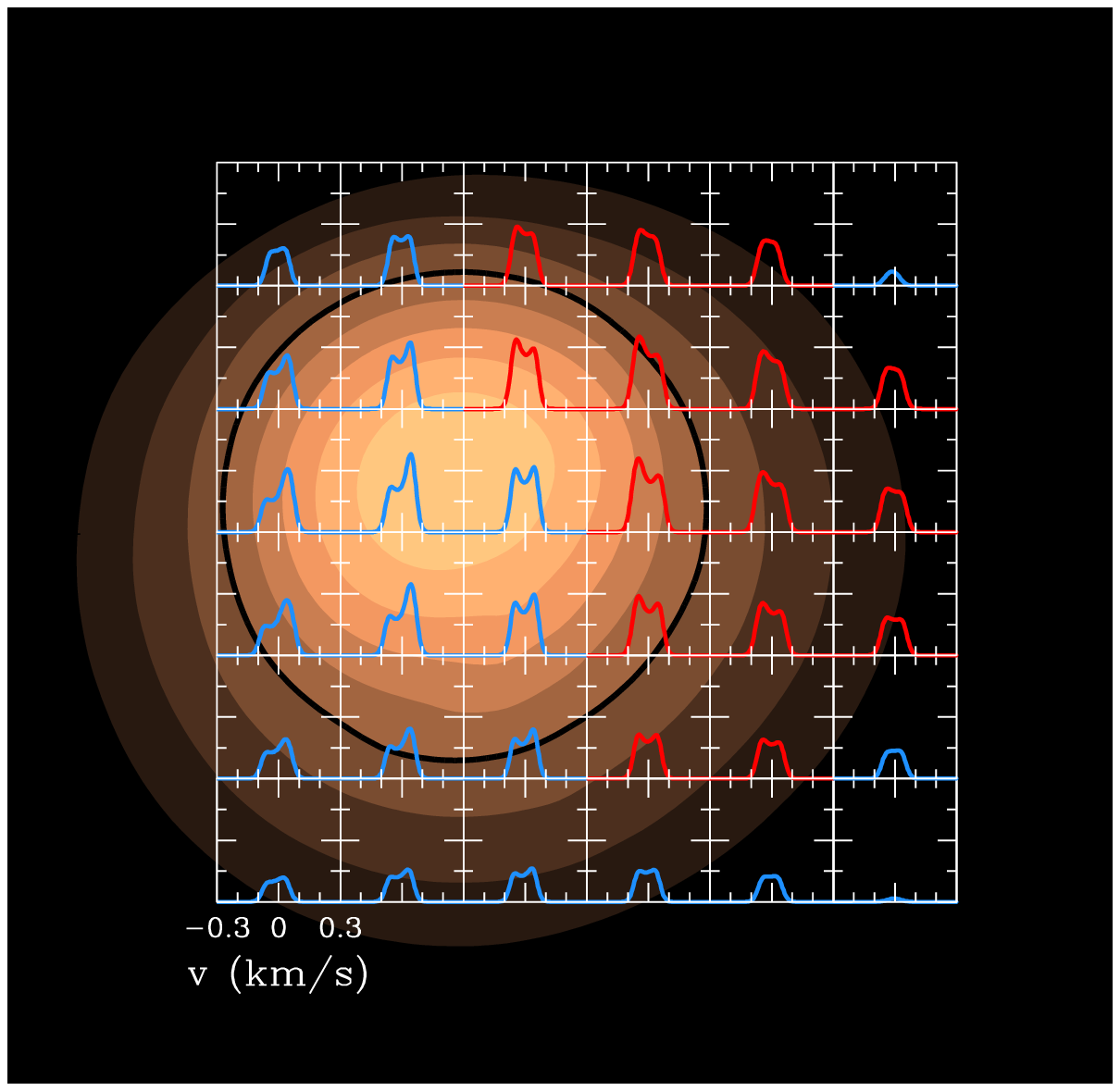}
\end{center}
\caption{A map of the CS(2-1) line profile overlaid upon the gas column
  density for the dipolar oscillation after more than
  $5\times10^5\,\yr$.  This is viewed from the same angle as shown in
  Figure \ref{fig:qcds}, ($\lambda=30^\circ$, $\theta=30^\circ$).  The
  gas density is shown as a filled contour map, with contours equally
  spaced in $10\%$ increments of the maximum value and the $50\%$
  contour is marked by the thick black line.  The CS(2-1) line profiles
  are associated with lines of sight centered in each subplot.
  Blue (red)
  lines correspond to approaching (retreating) gas on the nearside of
  the cloud relative to the observer.
  Note the prominent single reversal, similar to that
  associated with cloud rotation.  For the dipole mode the distortion
  in the column density map is substantially less than that observed
  for the quadrupolar mode.}\label{fig:dcds} 
\end{figure}

\subsection{Hydrodynamic Evolution}
Because the thermal gas heating and cooling time, via collisional coupling
to dust at high densities \citep{Burke1983} and molecular line
radiation at low densities \citep{Goldsmith2001}, is short
in comparison to the typical oscillation period
(on the order of $10^5\,\yr$), the oscillations of dark-starless cores are well modeled
with an isothermal equation of state.
In particular, we use a
barotropic equation of state with an adiabatic index of unity, in
which case the isothermal evolution is also adiabatic.  Since the
unbounded isothermal gas sphere is unstable, we truncate the solution
at a given density contrast ($\rho(R)/\rho_c$), producing the
standard Bonnor-Ebert sphere solution.  This is done by inducing a phase
change in the equation of state, \ie,
\begin{equation}
P =
\left\{
\begin{aligned}
\rho \frac{kT}{\mu m_p} &&& \text{if } \rho\ge\rho(R)\,,\\
\left[ \frac{\rho}{\rho(R)} \right]^\epsilon \rho(R) \frac{k T}{\mu m_p} &&& \text{if } \rho(R) > \rho \ge \delta\rho(R)\frac{T}{T_{\rm ext}}\,,\\
\rho \frac{kT_{\rm ext}}{\mu m_p} &&& \text{otherwise}\,,
\end{aligned}
\right.
\end{equation}
where the intermediate state, with $\delta=0.8$ and $\epsilon =
\log(\delta)/\log(\delta T/T_{\rm ext}) \ll 1$ chosen to make $P(\rho)$
continuous, is introduced to avoid numerical artifacts at the
surface, and $T_{\rm ext}\gg T$.  The resulting radial density and
pressure profiles are shown in Figure \ref{fig:profiles}.  Note that
there is a transition region, associated with the intermediate state in
the equation of state, which distributes the sudden drop in density
over many grid zones.  Despite this, the pressure is continuous
(necessarily) and nearly constant outside the surface of the
Bonnor-Ebert sphere.  We find that generally our results are
insensitive to the particular equation of state of the exterior hot
gas, given that it is sufficiently dilute.  This is appropriate for
globules like B68 which are known to be surrounded by hot, low density
gas.

We employ a three-dimensional,
self-gravitating hydrodynamics code to follow the long term mode
evolution.  Details regarding the numerical algorithm and specific
validation for oscillating gas spheres (in that case a white dwarf)
can be found in \citet{BroderickRathore2006}, and thus will only be briefly
summarized here. 

The code is a second order accurate (in space and time) Eulerian
finite-difference code, and has been demonstrated to have a low
diffusivity.  Because the equation of state is barotropic, we use the
gradient of the enthalpy instead of the pressure in the Euler equation
since this provides better stability in the unperturbed configuration
\citep{BroderickRathore2006}.  The Poisson equation is solved via
spectral methods, with boundary conditions set by a multipole
expansion of the matter on the computational domain.

In order to assess the suitability of the code for application to
isothermal gas spheres we tested the stability to monopolar
perturbations as a function of central-to-surface density
contrast, finding instability for values above $13.5$.  This is in approximate
agreement with the prediction of instability for density constrasts
above 14.3 \citep{FosterChevalier1993}.
The deviation from $14.3$ is likely
due to the effective atmosphere produced by the exterior hot matter,
and in particular the material in the transition region of the
equation of state.

Henceforth, the specific physical parameters of the unperturbed model that we
use, chosen to be roughly applicable to B68, are
$\rho_c = 1.185\times10^{-18}\,\g \cm^{-3}$, $\rho(R)/\rho_c = 12.5$, and
$T = 10\,\K$.  In all cases the simulations were performed at resolutions of
$64^3$, $128^3$ and $256^3$. Comparison of the latter two indicated
that the numerical solution had converged by these resolutions.  The
results presented in the following sections are from the simulations
at $256^3$.

The initial perturbed states are computed in the linear approximation
using the standard formalism of adiabatic \citep[which in this case is also
isothermal, ][]{Keto2006} stellar oscillations \citep{Cox1980}.
The initial conditions of the perturbed gas sphere are then given by
the sum of the equilibrium state and the linear perturbation.  We
employ the same definition for the dimensionless mode amplitude as
given in the Appendix of \citet{Keto2006}.

Throughout the evolution, the amplitude of a given oscillation mode,
denoted by its radial ($n$) and angular ($l$ \& $m$) quantum numbers,
may be estimated by the explicit integral:
\begin{equation}
A_{nlm} = \omega_{nlm}^{-1}\int \d^3\!x \,\rho\, \bmath{v}\cdot\bmath{\xi}_{nlm}^\dagger\,,
\label{eq:amp}
\end{equation}
where $\bmath{v}$ is the gas velocity, $\bmath{\xi}_{nlm}$ is the mode
displacement eigenfunction, $\omega_{nlm}$ is the mode frequency and
the dagger denotes hermitian conjugation.\footnote{For stellar
  oscillations the mode amplitudes may be determined from the density
  perturbation alone using the velocity potential.  However, when
  there is a non-vanishing surface density, as is the case for the
  Bonnor-Ebert sphere, this is not possible and the velocity field
  must be used.}

\subsection{Molecular Line Simulations}

Assuming that the gas contains the usual interstellar molecules, we simulate molecular line
observations of the hydrodynamic model by calculating the non-LTE (non-thermodynamic
equilibrium) emission of the
CS(2-1) transition using our three-dimensional
ALI (accelerated $\Lambda$ iteration) code \citep{Keto2004}.
The densities, velocities, and constant temperature are taken from the hydrodynamic simulation.
The line broadening is assumed purely thermal.


The abundance of CS is based on a simplified model of CO chemistry from \citet{KetoCaselli2008}.
We briefly describe the model here. The abundance
of CO in the gas is decreased by two effects, by photodissociation at the edge of the cloud and 
by freezing onto dust grains in the high density center. We calculate the reduction in the CO
abundance due to these two affects and assume that the same reduction affects the CS molecule.
We assume, as in \citet{Keto2006}, an undepleted abundance of CS of
$6\times 10^{-9}$.  This is consistent with \citet{Tafalla2002}, who
cite a range in values derived from their observations of $3 - 9 \times 10^{-9}$.

\citet{TielensHollenbach1985} suggest that at the edge of the cloud where photodissociation is
important, the dominant cycle for the formation and destruction of CO is
\begin{equation}
{\rm C}^+ \rightarrow {\rm CH}_2 \rightarrow {\rm CO} \rightarrow {\rm C} \rightarrow {\rm C}^+\,.
\end{equation}
The time scale for the formation of CH$_2$ by radiative association is longer than for the
formation of CO from CH$_2$ and O. Therefore we assume that the rate of formation of CO is
given by that of CH$_2$. The CO cycle may then be described by three rate equations
for the creation of C$^+$, C, and CO plus the conservation condition that the total amount of
carbon is contained in these three species. The rate equations for photodissociation
depend on the mean visual extinction at each point, which we calculate using the
machinery of the three-dimensional radiative transfer code.

The distribution of CO between the gas and grain phases is controlled by the rate of depletion from the
gas phase by freezing onto dust grains and the rate of
desorption off dust grains by cosmic rays. The steady state abundance of CO in the gas 
phase is then given by the ratio of the depletion time to the sum of the depletion and desorption
times. Because the depletion time scales inversely with the collision rate while the desorption time
scales inversely with the cosmic ray ionization rate, the depletion of CO is dependent only on
the gas density because the cosmic ray ionization rate is the same throughout the cloud.

\section{Quadrupole Perturbations} \label{QP}
The primary mode discussed in \citet{Keto2006} was the first
quadrupole p-mode, \ie, the $n=1$, $l=m=2$ oscillation mode.  In
order to reproduce the observed velocities it was necessary to
consider a dimensionless amplitude of $0.25$, which produced
near-vanishing densities in some portions of the cloud, and is thus
clearly well into the non-linear regime.  

The evolution of the dimensionless amplitude of the $(1,2,2)$ mode, as
defined by equation (\ref{eq:amp}), is shown as a function
of time in Figure \ref{fig:quad_amp}.  The decay is well approximated
by an exponential.  However, the strongly non-linear nature of the
decay is evident in the changing decay constant (resulting in a noticeable
bowing of the fitted line in Figure \ref{fig:quad_amp}), which yields an initial
e-folding time of approximately $2\,\Myr$, corresponding to roughly
$8$ oscillation periods ($16$ pattern periods).
Since the intrinsic dissipation in the code has previously been shown
to be small \citep{BroderickRathore2006}, the decay in the amplitude is
dominated by non-linear mode interactions, resulting in the transfer
of energy from large-scale, low-order modes to smaller-scale,
higher-order modes
\footnote{Generally the amplitude evolution may be expected to depend upon the
initial mode spectrum.  We investigated the significance of this
dependence by introducing higher order modes with randomly chosen
amplitudes and phases.  We found that as long as the
quadrupole mode remained dominant (in terms of $A_{nlm}$) the results
were unchanged.}.  However, mass loss driven by large velocity
perturbations and mode coupling in
the artificial atmosphere (see Figure \ref{fig:profiles}) also
contribute to the decay.  It is not clear if these processes are
artifacts of our numerical simulation and the manner in which we
initially prepare the cloud or if they may be found in nature as well.
Thus our estimate of the lifetime should be considered a
{\em lower limit}.

The only two oscillations of comparable wavelength (less than
approximately $R/5$) that reach significant amplitudes are the $(0,2,2)$
and $(0,4,4)$ modes, whose couplings are likely enhanced by the Cartesian
geometry of the underlying computational grid.  In particular, there
is no evidence for an inverse-cascade, implying that the gas cloud
will not fragment.  Indeed, this can be explicitly seen in the density
and velocity structures in the equatorial plane (see Figure \ref{fig:q})
in which the mode energy is primarily dissipated into small scale
motions. Thus even vigorous oscillations (at least up to amplitudes of
25\%) do not appear to be sufficient to split the cloud into a
binary.

Despite the decay in amplitude, the quadrupole mode structure is
still apparent in the density distribution of the cloud after a number
of oscillation periods.  This is also apparent in the substantially
distorted H$_2$ column densities, shown in Figure \ref{fig:qcds} for a
viewing angle that reproduces the generic features of B68
($\theta$ and $\lambda$ corresponding to the lattitude and longitude,
respectively, of the line of sight and oriented such that
$\lambda=0^\circ$/$90^\circ$ corresponding to a view from the
bottom/left of the page in Figure \ref{fig:q}). Even after
many oscillations the column density maps are generally qualitatively
similar to observations of dark cores which often show clouds with an
aspect ratio of approximately 2:1
\citep{Ward-ThompsonMotteAndre1999, Bacmann2000}.

Similarly, the complex velocity patterns, as inferred from the shape
of the CS(2-1) line profiles, are present after multiple oscillations.
To the left and below the plots of the density and velocity
structure in Figure \ref{fig:q}, the CS(2-1) line profiles are shown
for a number of lines of sight.  Comparison of the early time and late
time line profiles suggests that the small scale motions, the result
of damping of the initial oscillation, do not significantly alter the
line shapes or, as a direct consequence, the inferred line-of-sight
velocity distribution throughout the cloud.  Due to the intrinsically
complex nature of the oscillation velocity field, the inferred
velocity map can mimic expansion, contraction, rotation, or a combination of
these.  For example, the velocity patterns inferred by the line shapes
shown in Figure \ref{fig:q} appear to correspond to rotation, while if
these were rotated by $45^\circ$ (diagonal lines of sight) they would
appear to correspond to expansion or contraction.

The structure of the inferred velocity map can become substantially
more complex as we approach the pole (the polar angle $\theta$
approaches $0^\circ$).  As seen in Figure \ref{fig:qcds}, for
$\theta\simeq30^\circ$ and azimuthal angle $\lambda\simeq30^\circ$
a quadrupolar oscillation can reproduce the general features of B68.
This includes a double reversal of the line-of-sight velocity across
the cloud (represented by the change from red to blue to red in the
line profiles) and the significantly distorted gas column density.  In
fact, for the view shown, the relative orientation of the column
density distortion and the velocity double reversal is similar to that
observed in B68!  This continues to be the case even after $5\times10^5\,\yr$.

As anticipated in \citet{Keto2006}, the qualitative
features of the linear column density and velocity maps are still
present in the non-linear computation.  In particular, both the
anisotropic nature of the column densities and the complex velocity
structure persist for a number of oscillation periods.
Nevertheless, the non-linear decay of the amplitude, and subsequent
small scale motions, does become significant on timescales of a $\Myr$
as evidenced by the decay of the mode amplitude in Figure
\ref{fig:quad_amp}.  As a consequence, non-linear mode damping
appears to be the dominant constraint upon whether such an oscillation
can be clearly observed in velocity maps.

\section{Dipole Perturbations} \label{DP}
Although not explicitly considered in \citet{Keto2006}, dipole
oscillations are a possible alternative interpretation to rotation,
and perhaps also capable of explaining the velocity structure of B68.
For this reason we considered the first dipole p-mode, \ie, the
$n=l=m=1$ oscillation mode, as well.

As a consequence of its longer wavelength, the amplitude of the
dipole mode must be larger than that of the quadrupole oscillation
mode to produce similar velocities.  Hence, we chose a dimensionless
amplitude of $0.5$.  Similar to the quadrupole case, this is also
likely to be strongly non-linear.

As in the quadrupole case, the dipole oscillation decays
primarily into small scale motions.  In both the quadrupole and dipole
case the cloud sheds a few percent of its mass from the outer layers,
contributing to the decay of the oscillation.  In the dipole case this
also produces a notable shift in the center of mass of the cloud as a
result of the asymmetric mass loss through the boundary of the
computational domain.

As with the quadrupolar oscillations the gas column densities are
initially strongly distorted.  However, in contrast to the quadrupole
mode, after the first oscillation period (approximately
$1.5\times10^5\,\yr$) they have become roughly circular (see, \eg, Figure
\ref{fig:dcds}).  This suggests that large amplitude dipole modes
alone are insufficient to generate the observed anisotropy in the
shapes of dark cores.

Figure \ref{fig:dip_amp} shows the evolution of the dimensionless
amplitude for the first dipole p-mode.  Again the evolving decay
constant implies that this is a strongly non-linear process.  However,
in this case no oscillations with similar amplitudes develop within
the duration of the simulation (the largest being the $(0,2,2)$ mode,
which again is likely due to the Cartesian nature of the grid).  The
mode lifetime is on the order of $2\,\Myr$, considerably longer than
the mode oscillation period.  As before, there is no indication of
cloud fragmentation.

Nevertheless, the velocity maps inferred from the CS (2-1)
line shapes, shown in Figures \ref{fig:d}, retain their dipolar
structure, even after many oscillations.  Note that
this implies that inferring the global dynamical state of the cloud
from line profiles alone is significantly complicated by the possible
existence of large-scale oscillations.  As shown in Figures
\ref{fig:d} \& \ref{fig:dcds}, for a variety of viewing angles, the
motion appears to mimic rotation, expansion or contraction, all while
in the same dynamical state.  This is still true after the column
density maps become nearly spherical.

\section{Conclusions} \label{C}
Despite their strongly non-linear nature, large-amplitude oscillations
of pressure-supported molecular clouds reproduce many of the
qualitative features of observations of starless cores.  These
include the complex velocity structures and highly distorted column
densities found in B68 and L1544, as well as mimicking rotation,
expansion, and contraction depending upon the mode structure and
viewing angle.  This may complicate efforts to discern the global
dynamical state of these objects from observations of molecular
line shapes alone.

However, it may be possible to distinguish oscillations from collapse
by comparing the spectral line profiles and velocities of volatile
molecules such as N2H+ and NH3 against those of more refractory
species such as CO, HCO+, or CS \citep{KetoField2005}.  In a simple
breathing mode oscillation, the gas velocities are highest towards the
outer regions of the cloud where the abundance of the carbon species
is not depleted by freeze-out.  In contrast, in the gravitational
collapse of unstable BE spheres, the gas velocities are highest in the
center of the cloud where the emission from the non-depleting nitrogen
molecules is highest.

We note that we have restricted our attention to hydrodynamic
oscillations of pressure-bounded, thermally-supported
spheres. This description seems appropriate for globules such as
B68 that are surrounded by hotter gas and have
density profiles that closely match those of Bonnor-Ebert
spheres. Whether the Bonnor-Ebert description is appropriate for
embedded cores remains to be determined. For example, if magnetic
stresses contribute significantly to cloud support, then additional
classes of oscillations exist, associated with the magnetic field
\citep[see,\eg,][]{Hennebelle2003,Galli2005}.

The lifetimes of hydrodynamic oscillations of starless cores like
B68 are sufficiently long (few $10^6$ yr) to be a significant
fraction of the expected lifetime of a molecular cloud ($10^7$
yr).  The oscillation lifetimes are dominantly limited by non-linear
mode-mode coupling.  Nevertheless, large amplitude oscillations
persist for many periods (and thus many sound crossing times), and
are therefore not expected to be particularly rare.  There are no
signs of cloud fragmentation, and thus amplitudes comparable to those
discussed here are insufficient to produce binaries without the
inclusion of additional physical processes.

\acknowledgments
A.E.B. gratefully acknowledges the support of an ITC Fellowship from
Harvard College Observatory.


\begin{thebibliography}

\bibitem[Alves, Lada, \& Lada(2001)]{AlvesLadaLada2001}
        Alves, J., Lada, C., \& Lada, E. 2001, \nat, 409, 159

\bibitem[Bacmann et al.~(2000)]{Bacmann2000}
        Bacmann, A., Andre, A.P., Puget, J.-L., Abergel, A., Bontemps, S. \& Ward-Thompson, D. 2000, \aap, 361, 555


\bibitem[Bergin et al.~(2001)]{Bergin2001}
        Bergin, E., Ciardi, D., Lada, C., Alves, J., Lada, E., 2001, \apj, 557, 209

\bibitem[Bok (1948)]{Bok1948}
        Bok, B., 1948, {\it Centennial Symposia}, Harvard Observatory Monographs No. 7, Harvard-College Observatory, Cambridge                                                                               
\bibitem[Broderick \& Rathore (2006)]{BroderickRathore2006}
	Broderick, A.~E. and {Rathore}, Y., 2006, \mnras, 372, 923

\bibitem[Burke \& Hollenbach (1983)]{Burke1983}
        Burke, J.~R. \& Hollenbach, D.~J., 1983, \apj, 265, 223

\bibitem[Caselli et al.(2002)]{Caselli2002a}
        Caselli, P., Walmsley, C., Zucconi, A., Tafalla, M., Dore, L. \& Myers, P. 2002a, \apj, 565, 331

\bibitem[Cox(1980)]{Cox1980}
	Cox, J.P., 1980, Theory of Stellar Pulsations, Princeton University Press: Princeton

\bibitem[Foster \& Chevalier (1993)]{FosterChevalier1993}
        Foster, P. \& Chevalier, R., 1993, \apj, 416, 303

\bibitem[Galli (2005)]{Galli2005}
        Galli, D., 2005, \mnras, 359, 1083

\bibitem[Goldsmith (2001)]{Goldsmith2001}
        Goldsmith, P., 2001, \apj, 557, 736

\bibitem[Hennebelle (2003)]{Hennebelle2003}
        Hennebelle, P., 2003, \aap, 411, 9

\bibitem[Keto et al.(2004)]{Keto2004}
        Keto, E., Rybicki, G., Bergin, E., Plume, R., 2004, \apj, 613, 355

\bibitem[Keto \& Field(2005)]{KetoField2005}
        Keto, E. \& Field, G., 2005, \apj, 635, 1151 

\bibitem[Keto et al.(2006)]{Keto2006}
	Keto, E., Broderick, A., Lada, C.J. \& Narayan, R., 2006, \apj, 652, 1366

\bibitem[Keto \& Caselli(2008)]{KetoCaselli2008}
	Keto, E. \& Caselli, P., 2008, {\em in preparation}

\bibitem[Lada et al.(2003)]{Lada2003}
        Lada, C., Bergin, E., Alves, J., Huard, T., 2003, \apj, 586, 286

\bibitem[Tafalla et al.~(2002)]{Tafalla2002}
        Tafalla, M., Myers, P., Caselli, P., Walmsley, C. \& Comito, C., 2002, \apj, 569, 815

\bibitem[Tielens \& Hollenbach(1985)]{TielensHollenbach1985}
	Tielens, A.G.G.M. \& Hollenbach, D., 1985, \apj, 291, 722

\bibitem[Ward-Thompson, Motte, \& Andre(1999)]{Ward-ThompsonMotteAndre1999}
        Ward-Thompson, D., Motte, F. \& Andre, P. 1999, \mnras, 305, 143


\end{thebibliography}
\end{document}